\documentstyle[12pt,aaspp4]{article}
\lefthead{Simpson \& Meadows}
\righthead{Nuclear spectrum of Cen~A}

\begin{document}

\title{The nuclear spectrum of the radio galaxy NGC~5128 (Centaurus~A)}

\author{Chris Simpson\altaffilmark{1} and Vikki Meadows}
\affil{Jet Propulsion Laboratory, California Institute of Technology,
4800 Oak Grove Drive, Pasadena, CA 91109}

\altaffiltext{1}{Present address: Subaru Telescope, National Astronomical
Observatory of Japan, 650 N.~A`Oh\={o}k\={u} Pl., Hilo, HI 96720}

\begin{abstract}
We present near-infrared spectra of the nuclear disk in the nearby
radio galaxy NGC~5128 (Centaurus~A). On the basis of the observed
strengths of the [\ion{S}{3}]~$\lambda$0.9532\,\micron\ and
[\ion{Fe}{2}]~$\lambda$1.2567\,\micron\ lines, we classify NGC~5128 as
a LINER. Modeling of the strengths of these and additional lines
suggests that the nuclear region is powered by shocks rather than
photoionization.
\end{abstract}

\section{Introduction}

The galaxy NGC~5128 hosts the extragalactic radio source
Centaurus~A. It is our nearest radio galaxy, and has been the subject
of intense study which has revealed a jet seen at radio,
near-infrared, and X-ray wavelengths (Meier et al.\ 1989; Joy et al.\
1991; Schreier et al.\ 1979), and an unresolved nucleus which appears
longward of 2\,\micron\ (Turner et al.\ 1992). Perhaps the most
obvious feature of NGC~5128, however, is the dust lane which straddles
the galaxy. Such prominent and well-ordered dust obscuration is more
commonly associated with spirals, and is unusual for an early-type
galaxy (although perhaps not for a radio galaxy: de Koff et al.\
1996). It is widely believed that the dust lane was formed from
material stripped from a dusty disk galaxy (Baade \& Minkowski 1954;
Thomson 1992), and a natural consequence of this merger would be
substantial star formation, which is indeed inferred (Quillen, Graham
\& Frogel 1993; Storchi-Bergmann et al.\ 1997).

The type of spectroscopic study normally performed on the {\em
nuclei\/} of active galaxies, however, has yet to be undertaken on
NGC~5128. Since the nucleus is not visible in ground-based optical
images, no astrometry has been performed to allow it to be acquired
for spectroscopy. Phillips (1982) assumed the nucleus was located at
the near-infrared `hot spot' of Kunkel \& Bradt (1971), a feature
which can now be identified with the infrared jet, and
Storchi-Bergmann et al.\ (1997) used a wide (10\arcsec) slit which
contained a significant contribution from material excited by the
young stars in the larger-scale disk. Recently, Schreier et al.\
(1998) have shown that there is a physically distinct
Pa$\alpha$-emitting region at the nucleus approximately 2\arcsec\ in
size, which they interpret as a disk. As similar, albeit larger, disks
have been seen in a number of FR\,I radio galaxies, a nuclear spectrum
might reveal whether this smaller-scale disk shares their properties.

The only spectrum of the nucleus alone is that of Meadows \& Allen
(1992, MA92), who used the Infrared Imaging Spectrometer on the 3.9-m
Anglo-Australian Telescope, IRIS (Allen et al.\ 1993), which has the
capability to image a field before inserting the dispersing
element. In this paper, we present further spectra taken with IRIS,
covering the wavelength range $0.9\,\micron < \lambda < 2.5\,\micron$,
and use them to investigate the excitation mechanism operating at the
nucleus of NGC~5128.

Throughout this paper, we adopt a distance of 3.6\,Mpc to NGC~5128
(Soria et al.\ 1996), corresponding to a projected linear scale of
17\,pc\,arcsec$^{-1}$.

\section{Observations and reduction}

Data were taken with both narrow (1\farcs4) and wide (5\farcs8) slits
at a spatial resolution of 0\farcs8\,pixel$^{-1}$. IRIS was configured
to use the {\it IJ\/} (0.9--1.5\,\micron) and {\it HK\/}
(1.4--2.5\,\micron) echelles and provided a spectral resolution of
$\lambda/\Delta\lambda \sim 400$ (narrow slit) or $\sim 100$ (wide
slit). The true nucleus of NGC~5128 was acquired by locating the
continuum peak at $K'$, and spectra were taken at either end of the
13\arcsec\ slit and differenced to remove the sky and underlying
galaxy. All data were taken with the slit oriented east--west in
approximately 2\arcsec\ seeing. Narrow slit {\it HK\/} echelle data
were taken on the nights of UT 1991 April 26/27 and 1991 June 25/26
for a total integration time of 7400\,s, and were presented in
MA92. Narrow slit {\it IJ\/} data totaling 2000\,s were taken on UT
1992 April 14. Wide slit data were taken on UT 1992 September 13, with
an exposure time of 1200\,s in each echelle.

Standard spectral reduction techniques were applied to remove the
spatial and spectral response of the detector, the curvature of the
echelle orders, and the apparent rotation of the slit, which changes
across the orders, before one-dimensional spectra were extracted. The
narrow slit spectra, which were taken in non-photometric conditions,
have been flux calibrated by scaling the continuum fluxes so that they
matched photometry from the images of MA92, which in turn were
calibrated using the aperture photometry of Turner et al.\ (1992). We
estimate that the systematic uncertainties introduced by this method
are less than 20\%; the ratios derived from line pairs within the {\it
IJ\/} or {\it HK\/} echelles are, of course, unaffected. Observations
of standard stars were used to calibrate the wide slit data.

\section{Results and analysis}

\begin{figure}
\plotone{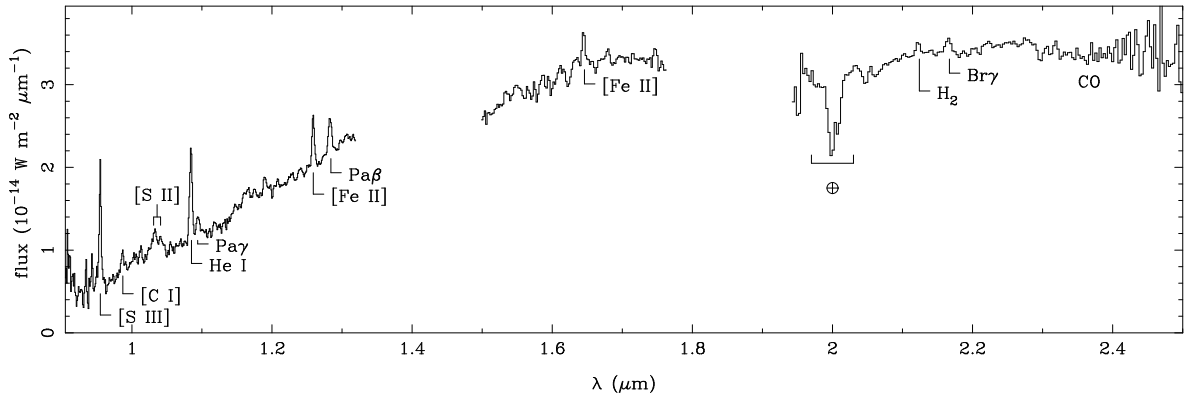}
\caption[]{Nuclear spectrum of NGC~5128, extracted through a $1\farcs4
\times 1\farcs6$ aperture. The emission lines are labeled, as are regions
of CO and atmospheric absorption. This spectrum was taken in
non-photometric conditions, and the flux scale has been bootstrapped as
described in the text.}
\label{fig:spek}
\end{figure}

The narrow slit spectrum is shown in Figure~\ref{fig:spek}. We present
the extracted line fluxes from both the wide and narrow slit spectra
in Table~\ref{tab:lines}. These fluxes were obtained by measuring the
signal above a linear continuum level which was determined by eye. The
placement of this continuum is the dominant source of uncertainty in
the measurements, and so several measurements with different
`reasonable' continuum levels were made to estimate the possible
errors. The measurement uncertainties from the narrow slit data are
much smaller than those from the wide slit spectra, due to the lower
sky noise, which allows more accurate continuum determination.

Given the poor seeing conditions of our observations, we would expect
the line fluxes from the wide slit spectra to be $\sim 40$\% higher
than those from the narrow slit spectra due to slit losses, even if
the line emission were unresolved. Coupled with the uncertainty in the
flux calibration of the narrow slit data, we are unable to find any
evidence for significant spatial extension in any of the emission
lines. It therefore seems likely that all the emission lines share the
morphology of Pa$\alpha$ seen by Schreier et al.\ (1998).

\subsection{Reddening to the emission line region}

The most reliable estimate of the reddening to the emission line
region comes from the near-infrared [\ion{Fe}{2}] lines, whose
intrinsic intensity ratio is fixed (Nussbaumber \& Storey 1988). The
observed ratio in the narrow slit data suggest $A_V = 3$\,mag,
although it is also consistent with zero reddening, and provides a
formal $3\sigma$ upper limit of $A_V < 8$. In the wide slit data, the
1.6435\,\micron\ line is strongly affected by poor sky subtraction and
so no useful ratio can be obtained.

The only reliable \ion{H}{1} line ratio we have from the wide slit
data is Pa$\beta$/Br$\gamma = 1.25 \pm 0.25$. Compared to the Case~B
ratio of 5.88, this implies an extinction of $A_V = 10.9 \pm 2.2$.
Although the Pa$\beta$/Br$\gamma$ and Pa$\gamma$/Br$\gamma$ ratios
from the narrow slit data are less reliable due to the method of flux
calibration, they too support a value for the extinction of about ten
magnitudes. However, the observed narrow slit Pa$\beta$/Pa$\gamma =
1.83 \pm 0.35$ is consistent with the Case~B value of 1.81, implying
negligible reddening, whereas if $A_V \approx 10$, it should be twice
as large. A formal $3\sigma$ upper limit to the extinction based on
this ratio (which is not affected by flux calibration uncertainties)
is $A_V < 7.7$\,mag. We therefore conclude that the Br$\gamma$ flux is
enhanced by an additional component which, presumably by virtue of
being heavily reddened, does not contribute to the Paschen lines. This
component could arise from a putative broad line region. Assuming the
broad line region suffers the $A_V \sim 30$\,mag of extinction which
obscures the nuclear continuum (Giles 1986; MA92) then any broad
Br$\gamma$ will be reduced in flux by a factor of about 20. If the
broad line is intrinsically 40 times brighter than the narrow line
(e.g.\ Jackson \& Eracleous 1995), then the ratio of narrow
Pa$\beta$/Br$\gamma \approx 4$, corresponding to $A_V \approx 3$, in
line with that derived from the [\ion{Fe}{2}] ratio. This extinction
is also consistent with obscuration from the dust lane alone, based on
the observed nuclear $J - H$ color if we assume that approximately
half the nuclear $H$ band emission is non-stellar (MA92), and so we
adopt $A_V = 3$\,mag for the extinction to the emission line region.
This value is somewhat lower than that derived by Schreier et al.\
(1996), but by failing to correct for the presence of the
heavily-reddened nucleus these authors are likely to have
overestimated the extinction to the NLR.

\subsection{Excitation mechanism}

MA92 suggested a starburst origin for the emission lines on the basis
of their {\it HK\/} spectrum, and in particular the
H$_2\,v$=1--0\,S(1)/Br$\gamma$ ratio. However, the low dissociation
temperature of H$_2$ means the H$_2$ line emission is not produced
cospatially with the other emission lines, and this line ratio is
therefore not ideal for determining the excitation mechanism. The
observed ratio of 0.6 is equally consistent with starburst, Seyfert,
and LINER galaxies (Moorwood \& Oliva 1988; Larkin et al.\ 1998). A
starburst can, however, be excluded on the basis of the luminosity of
the [\ion{Fe}{2}] emission. Following the discussion of van der Werf
et al.\ (1993), the radius of an individual SNR at the beginning of
the radiative phase is $\sim 6$\,pc, and our spectroscopic
observations indicate that the [\ion{Fe}{2}] emission is at best
barely resolved. Even making the generous assumption that the
line-emitting region is spherical with a diameter of 20\,pc, there is
only room for at most 5 SNRs but, adopting $L_{\rm[Fe II]} =
10^3\,L_\odot$ for a single remnant, about 20 are needed to provide
the extinction-corrected luminosity.

\begin{figure}
\plotone{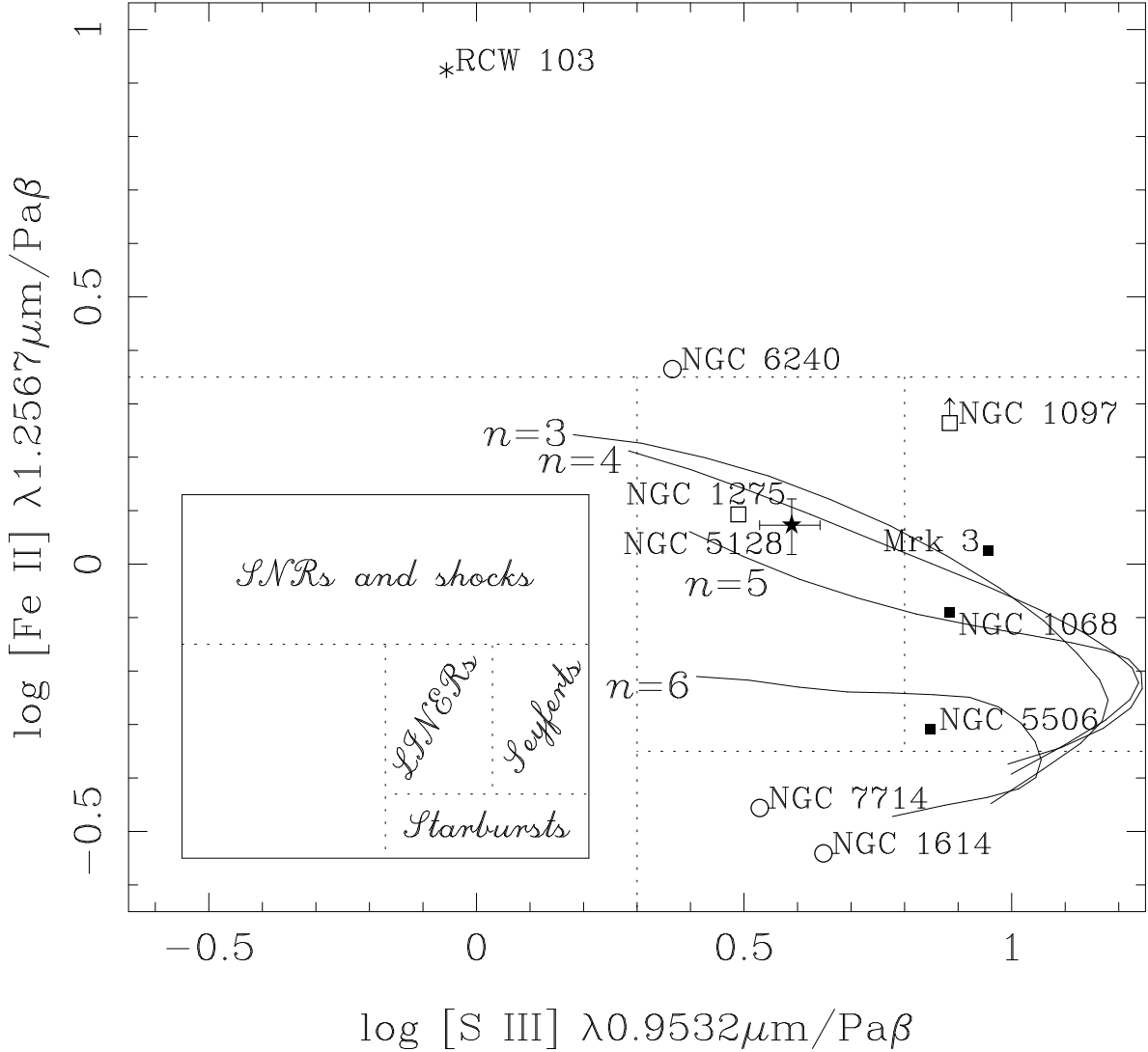}
\caption[]{[\ion{Fe}{2}]/Pa$\beta$ {\em vs\/} [\ion{S}{3}]/Pa$\beta$ line
ratio diagram. The regions expected to be occupied by different
classes of object are separated by dotted lines as shown in the inset,
and those few objects for which data exist for both forbidden lines
are plotted. We have plotted reddening-corrected line ratios for
Seyfert galaxies (solid squares), starburst galaxies (open circles),
and LINERs (open squares), using data from Kirhakos \& Phillips
(1989), Simpson et al.\ (1996), and Alonso-Herrero et al.\ (1997) and
references therein, assuming Case~B ratios to convert H$\alpha$ and
Br$\gamma$ fluxes to Pa$\beta$. The six-pointed star is the Galactic
supernova remnant RCW~103 (Dennefeld 1986). The solid lines are
theoretical predictions of photoionization models of different
densities, as described in the text.}
\label{fig:ratios}
\end{figure}

Of the emission lines we observe, those of [\ion{Fe}{2}] and
[\ion{S}{3}] have been the most well-studied, and we present a line
ratio diagram in Figure~\ref{fig:ratios}. Although few objects have
been observed in both emission lines, resulting in a sparsely
populated plot, many more have fluxes for one or other of the two
lines, and different classes of object are fairly well-separated in
each of the ratios, although there are regions of overlap. We can
therefore subdivide the parameter space as illustrated. We also plot
the theoretical line ratios as determined using the photoionization
code {\sc cloudy} V90.04 (Ferland 1996) for a plane-parallel slab with
densities in the range $n_{\rm H} = 10^3$--$10^6$\,cm$^{-3}$
illuminated by a power law spectrum extending from 10\,\micron\ to
50\,keV with $S_\nu \propto \nu^{-1.4}$, a range of ionization
parameter $U = 10^{-2}$--$10^{-5}$ and a gas phase iron abundance
one-twentieth of solar. The observed ratios can be explained by an
ionization parameter $U \sim 10^{-4}$, typical of LINERs, (e.g.\
Ferland \& Netzer 1983), and a density $n_{\rm H} \sim
10^{4.5}$\,cm$^{-3}$. This relatively high density is required to
produce the \ion{H}{1} line emission, since the extinction-corrected
H$\alpha$ luminosity is $L_{\rm H\alpha} = (2.2 \pm 0.3) \times
10^{32}$\,W (under Case~B), which is larger than the luminosity in the
nuclear disk of M~87 (Ford et al.\ 1994), yet the emission in NGC~5128
comes from a much smaller region.

We observe \ion{He}{1}~$\lambda$1.0830\,\micron/Pa$\gamma \approx 4$.
This compares with typical values of $\sim 2$ (from Case~B
recombination and the observed
\ion{He}{1}~$\lambda$0.4471\,\micron/H$\beta$ ratio) in both
\ion{H}{2} regions (Doyon, Puxley \& Joseph 1992) and AGN (Ferland \&
Osterbrock 1986), although larger ratios of 6.2 and 6.7 have been observed
in Orion and NGC~4151, respectively (Osterbrock, Shaw \& Veilleux 1990),
and a ratio of 5.8 is predicted for the starburst galaxy NGC~3256 based on
the observed \ion{He}{1}~$\lambda$2.06\,\micron/Br$\gamma$ ratio (Doyon,
Joseph \& Wright 1994). This ratio is therefore unable to tell us
anything about the excitation mechanism.

We have also detected the
[\ion{C}{1}]~$\lambda\lambda$0.9840,0.9850\,\micron\ and
[\ion{S}{2}]~$\lambda$1.0330\,\micron\ blends. They have also been
seen in NGC~4151, where their fluxes relative to
[\ion{S}{3}]~$\lambda$0.9532\,\micron\ were 0.01 and 0.05,
respectively (Osterbrock et al.\ 1990). We would expect stronger
emission from these low-ionization lines in a source with a lower
ionization parameter, but there are no data from such sources with
which to make comparisons, so we must compare our data with model
predictions. We again use {\sc cloudy} for the photoionization models
and {\sc mappings ii} (Sutherland \& Dopita 1993) to investigate the
effects of shock excitation, performing the same self-consistent
iterative process as Dopita \& Sutherland (1996). We used a solar iron
abundance for the shock models, since grains would be destroyed by the
shocks, liberating iron into the gas phase (Greenhouse et al.\
1991). A high preshock density is required, since the scaling relation
for H$\beta$ luminosity of Dopita \& Sutherland (1996) for a planar
shock with speed $V$ covering an area $d^2$, requires
\[
n \sim 1.1 \times 10^5 \left( \frac{V}{100\,{\rm km\,s}^{-1}}
\right)^{-2.41} \left( \frac{d}{10\,{\rm pc}} \right)^{-2} \, {\rm
cm}^{-3} \, ,
\]
to explain the inferred H$\beta$ luminosity.

Table~\ref{tab:models} presents the dereddened line fluxes and model
results (relative to Pa$\beta$). As {\sc mappings ii} does not provide
the strength of the \ion{He}{1}~$\lambda$1.0830\,\micron\ line in its
output, we are unable to compare the observed value with its
prediction. Instead we present a lower limit to the strength of this
line, from Case~B ratios and the strength of
\ion{He}{1}~$\lambda$0.4471\,\micron. The relatively high density
required will result in a somewhat larger flux, since
\ion{He}{1}~$\lambda$1.0830\,\micron\ is readily enhanced by
collisional excitation. Since all the emission lines we
are studying arise from a fairly small wavelength region, the line
ratios are not strongly affected by reddening. Even if the true
extinction were as high as $A_V = 5$\,mag, the most strongly affected
line ratio, [\ion{S}{3}]/Pa$\beta$, would only increase by 40\%, not
altering the conclusions we draw.

Although we have not investigated all the parameter space for the
shock models, due to the computational effort involved, it is clear
that shocks provide a rather better fit to the observed line ratios
than do the photoionization models. In particular, photoionization
fails to produce the observed line ratios over the range of ionization
observed (C$^0$, S$^+$, S$^{2+}$), as has often been seen in the
ultraviolet (e.g.\ Dopita et al.\ 1997). We therefore favor shocks
over photoionization, although ideally we would like some empirical
evidence to support this conclusion, from observations of the same
emission lines in objects where shocks are known to be the dominant
excitation mechanism.

\section{Discussion}

The above arguments support the classification of NGC~5128 as a LINER,
even though the emission lines which define this class (Heckman 1980)
are not observable. While Storchi-Bergmann et al.\ (1997) also made
such a classification, their much larger aperture was dominated by
emission lines originating in the starburst disk, and so did not
directly inform on the nuclear properties. The fact that we see a
shock-excited LINER spectrum in the nuclear disk is perhaps not
surprising, since the same is seen in M~87 (Virgo~A; Harms et al.\
1994; Dopita et al.\ 1997) and the similarities in the radio
properties of NGC~5128 and M~87 are quite extensive (Tingay et al.\
1998). Whether shock-excited nuclear disks are a common feature of
FR\,I radio galaxies is a question that requires further study.

Spectroscopic observations in the ultraviolet might also confirm our claim
that the emission lines are powered by shocks, since the predictions of
shock excitation and photoionization for lines such as
\ion{C}{4}~$\lambda$0.1550\,\micron\ and
[\ion{Ne}{5}]~$\lambda$0.3426\,\micron\ can differ by factors of
$10^4$ (e.g.\ Dopita et al.\ 1997). Although there is significant
extinction to the emission-line region, the emission lines are
sufficiently bright that these UV lines might still be
detectable. Indeed, our shock model, coupled with $A_V = 3$\,mag of
extinction, predicts $f_{\rm C\,IV} \approx 2 \times
10^{-18}$\,W\,m$^{-2}$, which is observable with the Space Telescope
Imaging Spectrograph (STIS). However, we note that the
\ion{C}{4} flux will be an order of magnitude fainter for every
magnitude of visual extinction by which we have underestimated the
true obscuration, so while its detection would unequivocally rule out
photoionization, its non-detection would not provide a definitive
result.

\section{Summary}

We have presented near-infrared spectra of the nucleus of NGC~5128,
covering the wavelength range 0.9--2.5\,\micron. The emission line
ratios are characteristic of a LINER, and can be modeled fairly well
by shock excitation or photoionization by a power law continuum
although shocks provide a noticeably better fit. Coupled with the
similarities between Cen~A and Vir~A, we favor shocks as the dominant
excitation mechanism in NGC~5128. Ultraviolet spectroscopy might be
able to confirm this through the detection of high-ionization lines
(e.g.\ \ion{C}{4}~$\lambda$0.1550\,\micron), but the high obscuration
in the UV could easily push them below a feasible detection threshold.

\acknowledgments

This work was performed at the Jet Propulsion Laboratory, California
Institute of Technology, under a contract with the National
Aeronautics and Space Adminstration. VM was supported for these
observations by an Australian Postgraduate Research Award. We thank
the Anglo-Australian Observatory for generous access to its computing
and other facilities. We also wish to thank David Allen for assistance
in acquiring the spectra and for many useful discussions, and the
anonymous referee for constructive suggestions.

\clearpage

\begin{table}
\caption[]{Emission line fluxes through the different spectroscopic
apertures. The $1\farcs4 \times 1\farcs6$ aperture measurement have been
calibrated as described in the text, and are subject to a $\sim 20$\%
systematic uncertainty, in addition to the random errors listed. Fluxes
indicated with a dagger ($\dagger$) are strongly affected by an uncertain
background level as a result of poor OH line subtraction.}
\label{tab:lines}
\begin{center}
\begin{tabular}{lccr@{ }c@{ }r@{}lr@{ }c@{ }r@{}l}
& $\lambda_{\rm rest}$ & \multicolumn{9}{c}{Line fluxes
($10^{-18}$\,W\,m$^{-2}$)} \\
\multicolumn{1}{c}{Line} & (\micron) & $1\farcs4 \times 1\farcs6$ &
\multicolumn{4}{c}{$5\farcs8 \times 1\farcs6$} &
\multicolumn{4}{c}{$5\farcs8 \times 2\farcs4$} \\
\hline
{}[\ion{S}{3}]      & 0.9532 & $54\pm4$ & 75&$\pm$& 7& &100&$\pm$&10& \\
{}[\ion{C}{1}]      & 0.9850 & $10\pm1$ & 17&$\pm$& 2& & 17&$\pm$& 3& \\
{}[\ion{S}{2}]      & 1.0330 & $40\pm5$ & 95&$\pm$& 9& &120&$\pm$&15& \\
\ion{He}{1}         & 1.0830 & $50\pm4$ &110&$\pm$&20&$^\dagger$ &
150&$\pm$&30&$^\dagger$ \\
Pa$\gamma$          & 1.0939 & $12\pm2$ & 33&$\pm$& 8&$^\dagger$ &
 48&$\pm$&12&$^\dagger$ \\
{}[\ion{Fe}{2}]     & 1.2567 & $26\pm2$ & 33&$\pm$& 3& & 32&$\pm$& 5& \\
Pa$\beta$           & 1.2819 & $22\pm2$ & 35&$\pm$& 5& & 40&$\pm$& 4& \\
{}[\ion{Fe}{2}]     & 1.6435 & $25\pm4$ & 80&$\pm$&20&$^\dagger$ & 
 79&$\pm$&13&$^\dagger$ \\
H$_2$               & 2.1213 & $11\pm2$ & 17&$\pm$& 2& & 27&$\pm$& 5& \\
Br$\gamma$          & 2.1657 & $17\pm3$ & 28&$\pm$& 4& &
\multicolumn{3}{c}{---}&$^\dagger$ \\
\hline
\end{tabular}
\end{center}
\end{table}

\begin{table}
\caption[]{Comparison of excitation models. All fluxes have been
normalized to Pa$\beta = 100$, and the observations have been corrected
for reddening. The photoionization model has $n_{\rm H} =
10^5\,$cm$^{-3}$ and $U = 10^{-4}$. The shock excitation model is for
a 120\,km\,s$^{-1}$ shock with precursor in a medium with density
$n_{\rm H} = 3 \times 10^4$\,cm$^{-3}$ and an equipartition magnetic
field.}
\label{tab:models}
\begin{center}
\begin{tabular}{lcr@{ }c@{ }rrr}
\multicolumn{1}{c}{Line} & $\lambda_{\rm rest}$ ($\mu$m) &
\multicolumn{3}{c}{Measured} & photo & shock \\
\hline
{}[\ion{S}{3}]  & 0.9532 & 393&$\pm$&29 &  849 & 390 \\
{}[\ion{C}{1}]  & 0.9850 &  68&$\pm$& 7 &   62 &  50 \\
{}[\ion{S}{2}]  & 1.0330 & 251&$\pm$&31 &  682 & 186 \\
\ion{He}{1}     & 1.0830 & 291&$\pm$&23 & 1291 & $>109$ \\
{}[\ion{Fe}{2}] & 1.2567 & 121&$\pm$& 9 &   77 &  94 \\
Pa$\beta$       & 1.2819 & 100&$\pm$& 9 &  100 & 100  \\
\hline
\end{tabular}
\end{center}
\end{table}

\clearpage

\end{document}